\documentclass[prl,10pt,twocolumn,superscriptaddress,showpacs,showkeys]{revtex4}
\usepackage{amsmath}
\usepackage{latexsym}
\usepackage{amssymb}
\usepackage{graphics,epstopdf}
\usepackage{graphicx}
\usepackage[colorlinks=true, citecolor=blue, urlcolor=blue ]{hyperref}
\usepackage{float}
\usepackage{graphicx}
\usepackage{amsfonts}

\begin{document}

\title{Sharing of Nonlocality of a single member of an Entangled Pair Is Not Possible by More Than Two Unbiased Observers on the other wing}

\author{Shiladitya Mal}
\email{shiladitya.27@gmail.com}

\author{A. S. Majumdar}
\email{archan@bose.res.in}
\affiliation{S. N. Bose National Centre for Basic Sciences, Block JD, Sector III, Salt Lake, Kolkata  700098, India}

\author{Dipankar Home}
\email{quantumhome80@gmail.com}
\affiliation{Center for Astroparticle Physics and Space Science, Bose Institute, Kolkata 700091, India}

\begin{abstract}
We address the recently posed question as to whether the nonlocality
of a single member of an entangled pair of spin $1/2$ particles can be shared
among multiple observers on the other wing who act sequentially and independently of each other~\citep{SGP}. We first show that the optimality
condition for the trade-off between information gain and disturbance in
the context of weak or non-ideal measurements emerges naturally when
one employs a one-parameter class of positive operator valued measures (POVMs). Using this formalism we then prove analytically that it is impossible to obtain violation of the Clauser-Horne-Shimony-Holt (CHSH) inequality by more than two Bobs in one of the two wings using unbiased input settings with an Alice in the other wing.
\end{abstract} 



\maketitle

\section{\textbf{I. Introduction}}

The quantum theory of measurement is  counter-classical in a sense that in order to obtain any statistical information about the state, some disturbance of the state becomes unavoidable. Measurement does not change the state only if the system is in an eigenstate of the observable to be measured.  A von Neumann type measurement~\citep{vn} dubbed as strong measurement induces strong collapse transforming the initial state of the system into one of the eigenstates of the measured observable. This type of measurement yields maximum information about the measured system. On the other hand, there exist measurement schemes such as
weak measurement~\citep{wk} which provide less information about the system while affecting it minimally, thus indicating a trade-off. 
A pertinent question in this context is to what extent one may control the state change due to measurement while obtaining some information
about it.
Recently, it has been shown that such a trade-off between the degree of disturbance and the amount of information gain about the system may
be optimized using suitably chosen pointer states by employing weak
measurements without post-selection~\citep{SGP}.
 
Another counter-classical feature of quantum theory is the presence
of nonlocal correlations in the measurement outcomes of two or more
parties sharing certain types of quantum states. This property of
quantum systems manifested by the violation of local realist, e.g.,
Bell-CHSH~\citep{Bell,chsh} type inequalities has been well studied in
in the literature in the context of multipartite or multilevel
states~\citep{bmulti}.  
However, a new fundamental question on the sharing of non-locality by multiple observers was posed recently~\citep{SGP}: Can  the nonlocality pertaining to a single member of an entangled pair of particles be shared among more than two independent observers who sequentially perform 
measurements on the other member of the entangled pair?  Note that the monogamy constraints~\cite{monog} on
entanglement and nonlocality do not apply in this scenario since
the condition of no-signalling is violated.
Though the observers who perform consecutive measurements are independent
of one another, the observer(s) who perform the prior measurement(s)
implicitly signal to the latter one(s) through the choice of their
measurement(s).  
 
The motivation of the present work is to investigate the above question of sharing of
nonlocality of an entangled pair of two particles by multiple observers.
 The experimental scenario considered here~\citep{SGP} is as follows. One of the particles of an entangled pair is measured
by a single observer Alice on one side. There exist multiple observers
(Bobs) on the other side who act sequentially and independently.
 Using weak measurements with optimized pointer settings it was 
 shown~\citep{SGP} that Bell-CHSH inequalities between Alice and
 an arbitrary number of sequential Bobs can be consecutively 
 violated in case of biased
 observation settings used by the various Bobs. In other words,  
 the protocol works when one of the inputs to the various independent observers occurs a lot more often than the other input. Though, in the unbiased input scenario numerical evidence indicated that 
 violation of the CHSH inequality is impossible by more than two
 Bobs, it was left as an open problem to show this analytically~\citep{SGP}.
 
In this paper we study this problem analytically using the framework of unsharp measurements~\citep{BL2} or POVMs with
a single parameter, based upon 
 the notion of generalised observables beyond the usual framework projective valued measures (PVM) or sharp measurements.
In the measurement  process after interaction of the physical system  with the apparatus the latter indicates a particular value corresponding
to  the former. This indication is modelled by means of pointer observable assuming an actual value corresponding to a value of the physical quantity of interest. Actual measurements in which the apparatus are represented by broad meter states, are seldom compatible with PVMs.
On the other hand, the generalised notion of POVMs turns out to be very helpful not only in explaining some of the conceptual problems of quantum theory, such as joint measurability of non-commutative 
observables~\citep{jointmeas}, but in also performing tasks such as  probing non-locality~\citep{nonloc} when projective measurement cannot. There are non-separable mixed states for which the Bell-CHSH
inequalities are violated not for the usual pairs of sharp spins but only for suitable families of unsharp observables. This is an illustration of the fact that optimisation of information gain in measurements can under
certain conditions only be achieved with POVMs but not with PVMs. A comprehensive introduction to
the topic of POVMs and their application in quantum foundations and experiments can be found in the monographs~\citep{BL2,BL3} and references therein.
 
Using the formalism of POVMs we show here  that unsharp observables
 characterized by a single unsharpness parameter saturate the optimal pointer condition with respect to the trade-off between disturbance and information gain, a condition that was earlier obtained using numerical
 optimization~\citep{SGP}. We then apply this formalism to the case of the problem of sharing nonlocality by multiple observers, as mentioned
  above. We prove analytically that more than two consecutive violations of the CHSH inequality are impossible in the unbiased scenario.
  The plan of the paper is as follows. In Section II we provide
   a brief discussion
on the quantum theory of measurement and POVMs. In Section III we show how the optimality
condition for the information gain versus 
disturbance trade-off emerges naturally within the unsharp 
measurement formalism. In Section IV we prove analytically that 
nonlocality of an entangled pair of spin $1/2$ particles cannot be 
shared between Alice and more than two Bobs. We conclude with a
brief summary in Section V.

\section{\textbf{II.  Quantum measurements}}

 The minimal content of the notion of measurement in quantum 
 mechanics~\citep{BL1} is
given by the probability reproducibility requirement, according to 
which a
particular measurement scheme qualifies as a measurement of a given observable $E$
if for all initial states of the system the associated probability distributions of
$E$ are reproduced in the resulting statistics of pointer readings.
The information available
by a given measurement depends on the statistical dependencies established
by the interaction between the system and the apparatus.
Let $S$ be the system with associated Hilbert space $\mathcal{H_{S}}$, 
and $\mathcal{A}$ be the measuring apparatus with Hilbert space $\mathcal{H_{A}}$.  The initial joint state of system and 
the apparatus is transformed unitarily during {\it pre-measurement}, and is given by
\begin{eqnarray}
\mathcal{U}(\rho_{S}\otimes\rho_{A})\mapsto \mathcal{U}\rho_{S}\otimes\rho_{A}\mathcal{U}^{*}.
\end{eqnarray}
An explicit construction of pre-measurement for discrete sharp
 observables has been known since the work of von Neumann~\citep{vn}.
 Any pre-measurement of an observable determines a state transformer on a measurable space $(\Omega,\mathcal{F})$, $\mathcal{I} : \mathcal{F} \mapsto \mathcal{L}(T(H_{s}))$ through the relation
\begin{eqnarray}
\mathcal{I}_{M}(X)(\rho):= tr[\mathbb{I}\otimes Z (\mathcal{U}\rho\otimes\rho_{A}\mathcal{U}^{*})\mathbb{I}\otimes Z]
\end{eqnarray}
where, $X\in\mathcal{F}$, and $\mathcal{L}(T(H_{s}))$ is the set of operators acting on a set of density states. The state transformer summarizes all the features of the pre-measurement. It recovers the observable via the relation
\begin{eqnarray}
tr[E(X)\rho]=tr[\mathcal{I}_{M}(X)\rho]   \forall X \in \mathcal{F}, \rho \in T(H_{S})
\end{eqnarray}
The state transformer for projective measurement of a discrete observable $A$ with eigenvalues $a_{i}$s is given by
\begin{eqnarray}
\mathcal{I}_{M}(\rho)=\sum_{a_{i}\in X}P_{i}\rho P_{i}.
\end{eqnarray}
For an observable $A=\sum a_{i}P_{i}$ with eigenvalue $a_{i}$ and eigenprojectors $P_{i}$, pre-measurement is given by
\begin{eqnarray}
\mathcal{U}(\varphi\otimes\phi)=\sum P_{i}\varphi\otimes\phi_{i}, \varphi\in\mathcal{H_{S}},\phi\in\mathcal{H_{A}}.
\end{eqnarray}
Let $\mathcal{Z}=\sum z_{i}\mathcal{Z}$ be an observable of apparatus $\mathcal{A}$, known as pointer observable. The reduced state of the apparatus is given by
$W(\varphi)=\sum_{a_{i}}p_{\varphi}^{A}(a_{i})P[\phi_{i}]$ 
(all the $P[\phi_{i}]$ are not necessarily mutually orthogonal) with the probability reproducibility condition given by
\begin{eqnarray}\label{pr}
p_{\varphi}^{A}(a_{i})=p_{W(\varphi)}^{Z}(z_{i})
\end{eqnarray}
where $p_{\varphi}^{A}(a_{i})$ is the probability distribution of outcomes of the observable $A$ and $p_{W(\varphi)}^{Z}(z_{i})$ is that of the pointer observable.
Eq.(\ref{pr}) implies that the outcome probabilities for observable 
$A$ are recovered as the distribution of the pointer values in the final apparatus state.  The emerging observable out of this measurement scheme is given by
\begin{eqnarray}
E_{i}=\sum p_{W(\varphi)}^{Z}(z_{i})P_{j}.
\end{eqnarray}\\


Now, following \citep{BL3} let us see how POVM emerges quite naturally in an actual measurement on a two level system. Let us take the system-apparatus coupling as
\begin{eqnarray}
\mathcal{U}=\exp^{i\lambda \sigma.{\hat{n}}\otimes \mathcal{P}}.
\end{eqnarray}
where $\mathcal{P}$ is the momentum operator and the pre-measurement is given by
\begin{eqnarray}
|\Psi\rangle =\mathcal{U}(\varphi\otimes\phi)=\sum P_{i}\varphi\otimes\exp^{i\lambda a_{i}\mathcal{P}}\phi \nonumber\\
=C_{+}\varphi_{+}\otimes\phi_{+}+C_{-}\varphi_{-}\otimes\phi_{-}.
\end{eqnarray}
Vectors $\exp^{i\lambda a_{i}\mathcal{P}}\phi$ or $\phi_{\pm}$ need not be mutually orthogonal.
Next, to describe registration of spots on the screen, the pointer observable is modelled by $P_{\pm}$, projectors onto the upper and lower half of the screen. For unsharp observables the state transformer is given by the generalised Luder transformer as
\begin{eqnarray}\label{ld}
\mathcal{I}_{M}(\rho)=\sum_{a_{i}\in X}\sqrt{E_{i}}\rho \sqrt{E_{i}}.
\end{eqnarray}
This measurement scheme yields
\begin{eqnarray}
<\Psi |\mathbb{I} \otimes P_{\pm}\Psi>= |C_{+}|^{2}\langle\phi_{+}|P_{+}\phi_{+}\rangle +|C_{-}|^{2}\langle\phi_{-}|P_{+}\phi_{-}\rangle \nonumber\\
:=\langle\varphi | E_{\pm}\varphi\rangle,
\end{eqnarray}
where the effects $E_{\pm}$ constitute the unsharp spin observables actually measured in this experiment, given by
\begin{eqnarray}
E_{\pm}=\langle\phi_{+} |P_{+}\phi_{+}\rangle P[\varphi_{+}]+\langle\phi_{-} |P_{-}\phi_{-}\rangle P[\varphi_{-}],
\end{eqnarray}
with $E_{+}+E_{-}=\mathbb{I}$, and $E_{\pm}^{2}\neq E_{\pm}$, i.e., $\langle\varphi_{+}|E_{+}\varphi_{+}\rangle,\langle\varphi_{-}|E_{-}\varphi_{-}\rangle\neq 0$ or $1$. If the center of mass of the wave-packets $\phi_{\pm}$ were well separated and localized in the appropriate half planes, i.e., if $\langle\phi_{\pm}|P_{\pm}\phi_{\pm}\rangle=1$, then $\langle\phi_{\pm}|P_{\mp}\phi_{\pm}\rangle =0$, in which case $E_{\pm}$ coincides with $P[\varphi_{\pm}]$. However, due to spreading of this wave-packet this coincidence is achieved only approximately. This provides a possible source of inaccuracy due to quantum indeterminism inherent in the center of mass wave-function. Even if spin is prepared sharply {\it a priori}, its value can only be ascribed with some uncertainty.

\section{\textbf{III. Optimality condition for information gain versus disturbance trade-off using unsharp measurements}}


Following the work of \citep{SGP}, let us
  consider a spin $1/2$ particle whose initial state is described by
  the state $|\psi\rangle (=\alpha|0\rangle +\beta |1\rangle)$. Considering von Neumann type measurements, after interaction with a meter with the state $\phi(q)$, the joint state of system and apparatus goes to $\alpha|0\rangle\otimes\phi(q-1)+\beta|1\rangle\otimes\phi(q+1)$. On tracing out the pointer state the reduced state of the system is given by 
\begin{eqnarray}\label{rs1}
\rho'=F\rho + (1-F)(\pi^{+}\rho\pi^{+}+\pi^{-}\rho\pi^{-}),
\end{eqnarray}  
 where $\rho=|\psi\rangle\langle\psi |,$ and $\pi^{\pm}$ are projectors onto states $|0\rangle ,|1\rangle$, and  $F(\phi)=\int_{-\infty}^{\infty}\langle\phi(q+1)|\phi(q-1)\rangle dq$, is called the quality factor of the measurement.  The probability of getting outcomes `$\pm$' is given by 
\begin{eqnarray} \label{pr1}
 p(\pm)=G\langle\psi|\pi^{\pm}|\psi\rangle +(1-G)\frac{1}{2}
\end{eqnarray} 
  Here $G=\int_{-1}^{1}\phi^{2}(q)dq$, which quantifies the precision of the measurement. It is independent of the state of the spin and depends on the width of the pointer compared to the distance between the eigenvalues. 
   These two parameters $F$ and $G$ characterize a weak measurement (the 
   case with $F=0$ and $G=1$ corresponds to a strong measurement).
  It was found in \citep{SGP} that a square pointer yields the relation $G=1-F$. However, such a pointer is not optimal. An optimal pointer is defined as the one which gives the best trade-off, i.e., for a given quality factor $F$, it provides the largest precision $G$. It was shown that the optimal 
information-disturbance trade-off condition given  by   
\begin{eqnarray}
F^{2}+G^{2}=1
\label{optcond}
\end{eqnarray}
emerges using an optimal pointer that is not a Gaussian wave-packet.

For two outcome measurements the notion of unsharp measurement discussed in section-II is
captured by the effect operator, $E^{\lambda}=(\mathbb{I}+\lambda n_{i}\sigma_{i})/2, i=1,2,3.$, with $\lambda\in(0,1]$. Thus, the set of effects can be written as a 
linear combination of sharp projectors with white noise, $E^{\lambda}\equiv \lbrace E_{+}^{\lambda},E_{-}^{\lambda}\vert E_{+}^{\lambda}+E_{-}^{\lambda}=\mathbb{I}\rbrace$, given by
\begin{eqnarray}
E_{\pm}^{\lambda}
=\lambda P_{\pm}+\frac{1-\lambda}{2}\mathbb{I}.
\end{eqnarray}\\
 In the unsharp formalism the non-selective un-normalised state of the system after pre-measurement according to the Luder transformation rule (\ref{ld}) is given by $\rho'=\sqrt{E_{+}^{\lambda}}\rho\sqrt{E_{+}^{\lambda}} + \sqrt{E_{-}^{\lambda}}\rho\sqrt{E_{-}^{\lambda}}$. From this we get
\begin{equation} \label{rs2} 
\rho'=\sqrt{1-\lambda^{2}}\rho + (1-\sqrt{1-\lambda^{2}})(P_{+}\rho P_{+}+P_{-}\rho P_{-}).
\end{equation}
The probabilities of getting the outcomes $\pm$ are given by
\begin{eqnarray}\label{pr2}
p(\pm)=tr[E_{\pm}^{\lambda}\rho]=\lambda tr[P_{\pm}\rho]+\frac{1-\lambda}{2}.
\end{eqnarray}
Comparing the two formalisms, i.e., comparing Eq.(\ref{rs1}) with Eq.(\ref{rs2})
 and Eq.(\ref{pr1}) with Eq.(\ref{pr2}), one sees that $G=\lambda$ and  $F=\sqrt{1-\lambda^{2}}$.
 Hence, $\lambda$ characterises the precision of the measurement. For $G=\lambda=1$, $F$ becomes zero,  this being the limit of sharp measurement.
 We thus find  that the optimal pointer state condition, $F^{2}+G^{2}=1$
 given by Eq.(\ref{optcond}) and derived through an optimization in \citep{SGP} emerges explicitly within the formalism of unsharp measurements. 
 In other words, unsharp measurement yields the maximum information about the system while disturbing the original state minimally.

\section{\textbf{IV. Sharing of non-locality}}

We now show that an application of the formalism of unsharp measurements in the 
context of sharing of non-locality enables us to resolve an open issue stated in \citep{SGP}. We show here analytically that using a pair of entangled spin $1/2$ particles Alice cannot share non-locality with more than two Bobs. All the observers have two measurement choices which they perform one at a time. Here it is important to note that each Bob measures independently of the previous Bobs on the particle of his possession. Hence, any Bob has to consider the average effect of possible choices of measurements done by previous Bobs~\citep{SGP}.

The joint probability of getting the outcome `$a$' and `$b_{n}$' by Alice and Bob${}^n$ (the $n$-th Bob) respectively, is given by
\begin{eqnarray}
p(a,b_{n})=p(a)p(b_{n}|a)=\frac{1}{2}Tr[\frac{\mathbb{I}+\lambda_{n}b_{n}\hat{y_{n}}.\vec{\sigma}}{2}\rho_{n|y_{1}...y_{n-1}}],
\end{eqnarray}
where $\rho_{n|y_{1}...y_{n-1}}$ is the state of the pair of spin-$\frac{1}{2}$ particles before the measurements of Alice and $Bob^{n}$, and $y_{i}$ is the measurement done by the $i$-th Bob.
For two Bob measuring in succession, the joint probability is given by
\begin{eqnarray}
p(a,b_{2})=\frac{\sqrt{1-\lambda_{1}^{2}}}{2}\frac{1-a b_{2}\lambda_{2}\hat{y_{2}}.\hat{x}}{2}\nonumber\\
+\frac{1-\sqrt{1-\lambda_{1}^{2}}}{2}\frac{1-a b_{2}\lambda_{2}\hat{x}.\hat{y_{1}}\hat{y_{1}}.\hat{y_{2}}}{2}
\end{eqnarray} 
The measurement directions chosen for Alice are $\hat{X},\hat{Z}$, and those for 
Bob are $\frac{-(\hat{Z}+\hat{X})}{\sqrt{2}},\frac{-\hat{Z}+\hat{X}}{\sqrt{2}}$. For the first Bob measuring weakly and the second Bob measuring sharply, the  CHSH values are given by $CHSH_{AB_{1}}=2\sqrt{2}\lambda_{1}$, and $CHSH_{AB_{2}}=\sqrt{2}(1+\sqrt{1-\lambda_{1}^{2}})$ respectively. This result coincides with that obtained in \citep{SGP}. In this case both Bobs obtain violation of the Bell-CHSH inequality when the precision of the $1$st Bob remains within the range $1/\sqrt{2}$ and $\sqrt{2(\sqrt{2}-1)}$.

Now consider the case of three Bobs with a single Alice. In this case the $1$st 
and $2$nd Bobs both measure weakly, while the last Bob measures sharply. We get
\begin{eqnarray}
p(a,b_{3})=\frac{1}{2}[\sqrt{1-\lambda_{1}^{2}}\sqrt{1-\lambda_{2}^{2}}\frac{1-a b_{3}\lambda_{3}\hat{y_{3}}.\hat{x}}{2}+\nonumber\\
(1-\sqrt{1-\lambda_{1}^{2}})\sqrt{1-\lambda_{2}^{2}}\frac{1-a b_{3}\lambda_{3}\hat{x}.\hat{y_{1}}\hat{y_{1}}.\hat{y_{3}}}{2}+\nonumber\\
\sqrt{1-\lambda_{1}^{2}}(1-\sqrt{1-\lambda_{2}^{2}})\frac{1-a b_{3}\lambda_{3}\hat{x}.\hat{y_{2}}\hat{y_{2}}.\hat{y_{3}}}{2}+\nonumber\\
(1-\sqrt{1-\lambda_{1}^{2}})(1-\sqrt{1-\lambda_{2}^{2}})\frac{1-a b_{3}\lambda_{3}\hat{x}.\hat{y_{1}}\hat{y_{1}}.\hat{y_{2}}\hat{y_{2}}.\hat{y_{3}}}{2}]
\end{eqnarray}
Here $\lambda_{2}$ is precision of measurement by the $2$nd Bob.
The correlation between Alice and Bob${}^3$ is given by
\begin{eqnarray}
C_{3}=\lambda_{3}[\sqrt{1-\lambda_{1}^{2}}\sqrt{1-\lambda_{2}^{2}}\hat{y_{3}}.\hat{x}+\nonumber\\
(1-\sqrt{1-\lambda_{1}^{2}})\sqrt{1-\lambda_{2}^{2}}\hat{x}.\hat{y_{1}}\hat{y_{1}}.\hat{y_{3}}+\nonumber\\
\sqrt{1-\lambda_{1}^{2}}(1-\sqrt{1-\lambda_{2}^{2}})\hat{x}.\hat{y_{2}}\hat{y_{2}}.\hat{y_{3}}+\nonumber\\
(1-\sqrt{1-\lambda_{1}^{2}})(1-\sqrt{1-\lambda_{2}^{2}})\hat{x}.\hat{y_{1}}\hat{y_{1}}.\hat{y_{2}}\hat{y_{2}}.\hat{y_{3}}].
\end{eqnarray}
As any Bob is ignorant about inputs of previous Bobs, this correlation has to be averaged over all possible earlier inputs. Hence, one has
\begin{eqnarray}
\bar{C_{3}}=\sum_{y_{1}y_{2}}C_{3}P(y_{1})P(y_{2})
\end{eqnarray}\label{chsh3} 
With this average correlation we find the CHSH sum between Alice and Bob${}^3$
given by
\begin{eqnarray}\label{I3}
\mathcal{I}^{3}=\frac{(1-\sqrt{1-\lambda_{1}^{2}})(1-\sqrt{1-\lambda_{2}^{2}})}{\sqrt{2}}
\end{eqnarray}
For the $1$st and $2$nd Bobs the corresponding CHSH values are given by $CHSH_{AB_{1}}=2\sqrt{2}\lambda_{1}$ and $CHSH_{AB_{2}}=\lambda_{2}\sqrt{2}(1+\sqrt{1-\lambda^{2}})$, respectively.
In order for the $1$st Bob to obtain violation of the Bell-CHSH inequality, his measurement precision $\lambda_{1}$  has to be greater than $1/\sqrt{2}$. For the $2$nd Bob to get the violation, it is required that $\lambda_{2}>\frac{2}{\sqrt{2}+1}$. Thus, it follows from Eq.(\ref{I3}) that if the first two Bobs 
obtain violation, the subsequent CHSH value corresponding to Bob${}^3$ cannot be
greater than $2$. In the worst case scenario of violation of CHSH by Bob${}^1$ and Bob${}^2$, i.e., when both of them obtain minimal violation, $\mathcal{I}^{3}$ can not becomes greater than $1.88$. It is thus clear that more than two Bobs can never share the 
nonlocality of a pair of spin $1/2$ particles with a single Alice, a result that
was numerically conjectured in \citep{SGP}. One may
note that the sequence of the particular Bobs is not important in this scenario.
For example, Bob${}^3$ may obtain violation if the sharpness of the $2$nd Bob's measurement is sufficiently weak for the latter not to get a violation. There 
exists a range of unsharpness parameters for each Bob so that any one pair of
Bobs in
the combinations (Bob${}^1$, Bob${}^2$),  (Bob${}^1$, Bob${}^3$), or (Bob${}^2$, Bob${}^3$) can simultaneously demonstrate non-locality.

\section{\textbf{V. Discussions}}

In this work we have considered the question of sharing the nonlocality of a single member of an 
entangled pair of spin $1/2$ particles by multiple observers on the other side acting sequentially
and independently of each other. This 
fundamental issue has
been recently studied by Silva et al.~\citep{SGP} within the context of the trade-off between information
gain and disturbance in weak measurements. It was found that 
multiple Bobs having sequential access to one particle from the entangled
pair can indeed violate Bell-CHSH inequalities with a single Alice on the 
other side, provided biased measurement settings were used.
It was also observed therein using numerical evidence that in the case of unbiased
input settings, it was not possible for more than two Bobs to obtain
CHSH violations, and it was left as an open problem to prove 
this analytically~\citep{SGP}. In the present work we address this issue by
considering usharp measurements within the framework of POVMs~\citep{BL2}.
We first show that the optimality condition for the information gain versus 
disturbance trade-off derived in \citep{SGP} emerges naturally using one-parameter POVMs. Applying this framework to the problem of sharing of nonlocality then
enables us to show analytically that more than two Bobs cannot violate the CHSH inequality with a single Alice.

{\it Acknowledgements}: ASM and DH acknowledge support from the project SR/S2/LOP-08/2013
of DST, India.


\end{document}